\documentclass[aps,prb,twocolumn,showpacs,superscriptaddress,floatfix]{revtex4-1}

\pdfoutput=1
\usepackage[titletoc,toc,title]{appendix}
\usepackage{amsmath,amssymb}
\usepackage{graphicx}
\usepackage{color}
\usepackage{rotating}
\usepackage{bm}
\usepackage{setspace}
\usepackage{hyperref}
\usepackage{float}
\usepackage{soul}
\usepackage[T1]{fontenc}
\usepackage{bbold}
\usepackage{color}
\usepackage{braket}

\hypersetup{
colorlinks=true,
urlcolor= blue,
citecolor=blue,
linkcolor= blue,
bookmarks=true,
bookmarksopen=false,
}

\renewcommand{\vec}[1]{\bm{#1}}

\begin{document}

\preprint{APS/123-QED}

\title{Electronic Structure Theory of Strained Two-Dimensional Materials with Hexagonal Symmetry}

\author{Shiang Fang}
\affiliation{Department of Physics, Harvard University, Cambridge, Massachusetts 02138, USA.}
\author{Stephen Carr}
\affiliation{Department of Physics, Harvard University, Cambridge, Massachusetts 02138, USA.}
\author{Miguel A. Cazalilla}
\affiliation{Department of Physics, National Tsing Hua University and National Center for Theoretical Sciences (NCTS), Hsinchu 30013, Taiwan}
\affiliation{Donostia International Physics Center (DIPC), Manuel de Lardizabal, 4. 20018, San Sebastian, Spain}
\author{Efthimios Kaxiras}
\affiliation{Department of Physics, Harvard University, Cambridge, Massachusetts 02138, USA.}
\affiliation{John A. Paulson School of Engineering and Applied Sciences, Harvard University, Cambridge, Massachusetts 02138, USA.}

\date{\today}

\begin{abstract}
We derive electronic tight-binding Hamiltonians for strained graphene, hexagonal boron nitride and transition metal dichalcogenides based on Wannier transformation of {\it ab initio} density functional theory calculations. Our microscopic models include strain effects to leading order that respect the hexagonal crystal symmetry and local crystal configuration, and are beyond the central force approximation which assumes only pair-wise distance dependence. Based on these models, we also derive and analyze the effective low-energy Hamiltonians. Our {\it ab initio} approaches complement the symmetry group representation construction for such effective low-energy Hamiltonians and provide the values of the coefficients for each symmetry-allowed term. These models are relevant for the design of electronic device applications, since they provide the framework for describing the coupling of electrons to other degrees of freedom including phonons, spin and the electromagnetic field. The models can also serve as the basis for exploring the physics of many-body systems of interesting quantum phases.
\end{abstract}

\maketitle



\section{\label{sec:level1}INTRODUCTION}

Strain effects are important in the physics of van der Waals two-dimensional materials\cite{graphene_rev,2dmaterial_rev0}, which feature covalent bonding within each single layer and weaker attraction between layers. Instead of being geometrically flat, these materials exhibit ripples and corrugations, features that are ubiquitously observed, for example, in free-standing graphene\cite{graphene_ripple} and in samples on a substrate\cite{graphene_ripple_sub}. After the discovery of graphene, the list of two dimensional materials has been constantly growing, and includes now several materials, such as hexagonal boron nitride (hBN)\cite{hBN}, black phosphorus\cite{black_phosphorus}, and transition metal dichalcogenides (TMDCs)\cite{2dmaterial_rev2} with chemical composition MX$_2$ (M= transition metal atoms Mo, W; and X= chalcogen atoms S, Se, Te.). These layered materials exhibit interesting behavior ranging from topological phases\cite{TMDC_TI} to superconductivity\cite{TMDC_sc}, magnetism\cite{TMDC_magnetism}, topological order and anyonic excitations in fractional quantum Hall liquids\cite{graphene_fqhe}, and other strongly correlated phases that arise due to the reduced dimensionality and screening\cite{TMDC_cdw}. The list of their possible applications is also constantly expanding, including devices for optoelectronics\cite{2dmaterial_rev2}, plasmonics\cite{tmdc_plasmon} and valleytronics\cite{valleytronics}, which involve structures based on single-layer or heterostructure form\cite{2dmaterial_rev1}. These stable layers can sustain a substantial amount of external strain, as high as 25\% in graphene\cite{graphene_strain_exp}. Kirigami structures based on graphene\cite{Kirigami_graphene} allow even higher degree of stretchability and resilience. Scanning tunneling microscopy~(STM)\cite{graphene_stm_strain} or atomic force microscopy~(AFM)\cite{graphene_afm_strain} tips can be used to introduce indentation and strain in a controlled manner. The strain-induced time-reversal symmetric pseudomagnetic field in graphene has been shown to reach 300T\cite{graphene_strain_LL}. A desirable functionality would be to use strain and deformation to manipulate the flow of electrons or excitons in the design of layered-material based devices\cite{funnel_1,funnel_2}, and the associated nanostructures such as nanoribbons\cite{GNR_Strain}. To achieve this goal, reliable quantitative understanding and modeling of the strained-layered properties are crucial and call for a more systematic treatment than what is presently available.

Conventional approaches for modeling can be classified in two categories: The top-down method treats the deformed layers as a manifold with curvature and local metric tensor structure, analogous to a membrane in soft matter\cite{membrane} and to general relativity in curved space-time\cite{curve_space}. In this approach, once the differential geometry tensors are constructed from the deformed layers, they couple to the low energy effective field theories as symmetry-allowed gauge fields, potentials and connections\cite{novel_strain_review,rev_graphene_gauge,graphene_curv,curve_space2,TMDC_curvature}. The bottom-up approach relies on computationally demanding first-principles calculations\cite{strain_tmdc} or on scaling of tight-binding matrix elements in the presence of the lattice deformation\cite{graphene_rev,graphene_unistrain_tb,tmdc_tb_strain}. The scaling of these coupling terms is usually parametrized empirically as a function of pair-wise distances, which is known as the central force approximation, in the form of Gr\"{u}neisen parameters\cite{tmdc_strain_curv}. In practice, these empirical parameters are usually obtained from fitting band structure calculations of the deformed crystal, which is relatively insensitive to the underlying orbital character and composition of the coupling terms. Potential pitfalls in this approach include overfitting of the band structure, distortions in the wavefunction character and the breakdown of the approximations invoked. Another issue arises from bridging the top-down and bottom-up approaches as pointed out by Yang\cite{boyang_graphene}: the proper "metric" and the emergent geometry in the low-energy model should stem from the deformation of the underlying tight-binding Hamiltonian, rather than being of purely geometric origin. It is thus valuable to derive from an {\it ab initio} perspective the tight-binding parameters of the strained layered crystals, especially for materials with multiple orbital symmetries and complicated character. Previously, we have demonstrated an efficient and reliable method for modeling layered materials and their vertical stacking\cite{tmdc_tbh,graphene_tbh} including intra- and inter-layer coupling terms based on the Wannier transformation of electronic band structures obtained from density functional theory (DFT) calculations, without having to rely on empirical fitting parameters. Here we generalize the Wannier method\cite{mlwf} to monolayers with in-plane strain and derive the relevant models, compatible with the underlying crystal symmetry. In increasing order of complexity with the underlying orbital content, we construct such Wannier tight-binding Hamiltonians (TBH) for graphene, hexagonal boron nitride (hBN) and four TMDCs. These models are valid in the presence of slowly-varying in-plane strain field, providing the electronic coupling to long-wavelength in-plane acoustic phonon modes\cite{graphene_raman,graphene_raman_eph}. We also derive the corresponding effective low-energy theories coupled with the strain field, consistent with the effective models derived from the principles of symmetry group representations, which by itself can identify all symmetry-allowed terms\cite{group_theory_md,eff_graphene_strain} but is insufficient to provide estimates for the values of the coupling constants involved. Our {\it ab initio} Wannier tight binding approach thus complements the powerful symmetry group analysis, gives accurate values of the parameters in the model, and empowers calculations of large-scale structures of strained materials\cite{straintronics,strain_2d_rev} and finite size system with coupling to external fields\cite{LL2D_Wannier}.

For the underpinning density functional theory calculations, we adopted the exchange correlation functional parametrized by Perdew, Burke and Ernzerhof (PBE)\cite{pbe}. Conventional DFT functionals tend to underestimate the band gap values derived from the experimental results. On the experimental side, various factors from the dielectric screening of the substrates\cite{Dielectric_genome} and doping\cite{dope_gap_renormalization} might further complicate the comparisons to theoretical band structure. In terms of the theoretical calculations, different choices of functionals such as HSE06\cite{MoS2_HSE} or more advanced GW calculations for quasi-particle energies\cite{TMDC_GW} can be adopted to improve the band gap values. In previous work, we surveyed briefly the comparison between theoretical calculations and experimental measurements\cite{strain_tmdc}. Here, we focus on the modeling of strain correction terms of the two-dimensional crystals with hexagonal symmetry. Further improvements of the electronic band structures from different choices of the functionals and more advanced GW calculations are compatible with the Wannier construction method\cite{tmdc_tbh} and the analysis presented here will apply with modified parameters.

The paper is organized as follows: In Sec. II, we first elaborate on the conventions of crystal structure and the assumptions involved in strain field modeling in our work. We then apply these methods to construct tight-binding Hamiltonians for the in-plane strained crystals in graphene, hBN and TMDCs. In Sec. III, we derive the effective low-energy Hamiltonians, based on the strained tight-binding Hamiltonians, and compare with symmetry group analysis. We conclude in Sec IV, which summarizes our work and points out the potential generalizations and applications of our models. We elaborate on the numerical framework for DFT calculations and Wannier constructions in Appendix A. In the Appendix B, we give the mathematical background of the symmetry group analysis and provide guidance for generalizing to other scenarios relevant to layered materials. The values of the tight-binding parameters including the effect of strain, for four TMDCs are also tabulated in the Appendix \footnote{The scripts for generating the TMDC Hamiltonians with strain will be available on the last author's research group website.}.


\section{\label{sec:level1}TIGHT-BINDING HAMILTONIAN FOR STRAINED LAYERED MATERIALS}

We develop the tight-binding Hamiltonians for the strained layered structures by following exactly the same procedure as in our earlier work for ideal layers\cite{tmdc_tbh,graphene_tbh} (see Appendix A for more detailed descriptions, and Fig. \ref{fig:wannier_construction} where the steps from DFT to Wannier model construction for the WSe$_2$ monolayer crystal are illustrated.). From the macroscopic point of view, the strain field is described within continuum elasticity theory. For strained layers, it is equally important to specify the underlying deformed microscopic configurations. Here, we provide the connection between the macroscopic elastic theory and the microscopic atomic details, using the generalized Cauchy-Born rule for the local optimum strain configuration of basis atoms, which might show violations in the restricted elastic relations. After establishing the deformed crystal structure, the tight-binding Wannier Hamiltonians of the relevant selected bands are constructed and truncated to retain only a few near neighbors, as appropriate for each layer type. These strain scaling parameters are tabulated along specific bond directions with simpler expressions, while other equivalent bonds are related by symmetry transformations.

\subsection{General Formulation of Strained Lattices}

\begin{figure}[h!]
\centering
\includegraphics[width=0.48\textwidth]{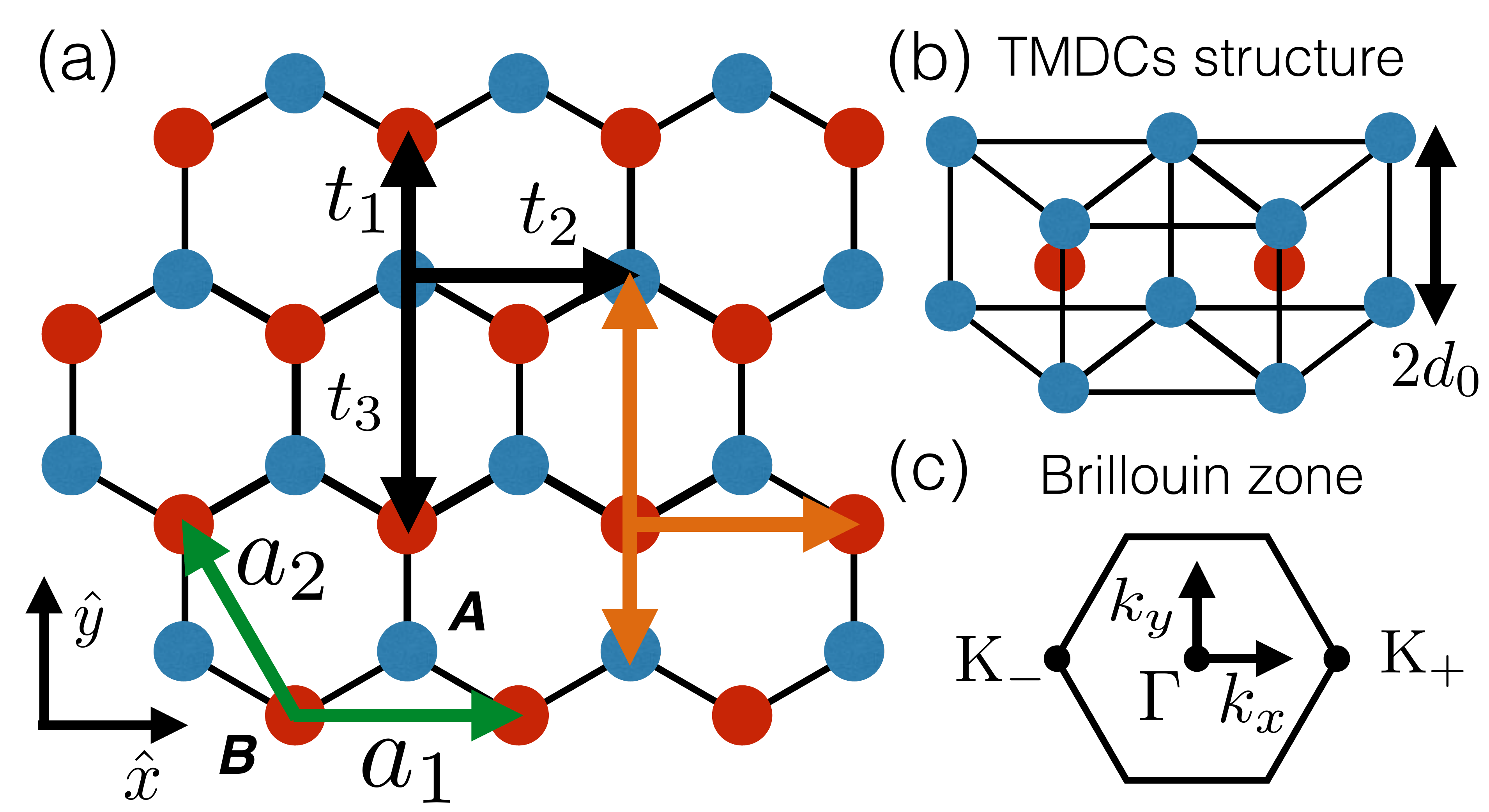}
\caption{The conventions for the honeycomb hexagonal crystal structure: (a) Top view of the crystal lattice with primitive vectors $\vec{a}_i$ with $A$ ($B$) basis atoms shown as blue (red) solid circles. In hBN, Boron (Nitrogen) atoms occupy sublattice $A$ ($B$) sites, while in TMDCs metal atoms (chalcogen pairs) sit at $B$ ($A$) sites. The three thick black arrows labeled by $t_i$ denote the hopping bonds used in strained graphene and hBN up to third nearest neighbors in Eq. (\ref{eqn:ghbn_hop}). For TMDCs, the hoppings from M sites are denoted by the thick orange arrows instead for Eq. (\ref{eqn:TMDC_first}) and (\ref{eqn:TMDC_second}). (b) Perspective side view of the trilayer structure in TMDC. (c) Brillouin zone in momentum space.}
\label{fig:crystal_convention}
\end{figure}

Graphene, hBN, and the TMDC layered materials investigated in our work share the hexagonal lattice system and the honeycomb crystal structure with periodic lattice vectors $\vec{a}_1=a\hat{x}$, $\vec{a}_2=a(\frac{-1}{2}\hat{x} +\frac{\sqrt{3}}{2}\hat{y})$ where $a$ is the lattice constant. Two basis sites are located in the projected layer plane $\vec{\delta}_B=\vec{0}$ and $\vec{\delta}_A=(2\vec{a}_1+\vec{a}_2)/3$. In hBN, the Nitrogen atom occupies the $\vec{\delta}_B$ site. The TMDC layer consists of three atomic sublayers in each single layer unit as shown in Fig. \ref{fig:crystal_convention} (b), with chalcogen atoms at projected sublattice sites $\vec{\delta}_A$ and at height $\pm d_0$ above and below the plane of the metal atoms. For the reciprocal space representation, these crystals share the Brillouin zone shown in Fig. \ref{fig:crystal_convention} (c), with special $k$ points K$_\pm=\pm \frac{4\pi}{3a}\hat{x}$ where the valleys appear in the band structure.

\begin{widetext}

\begin{figure} [h!]
\centering
\includegraphics[width=0.95\textwidth]{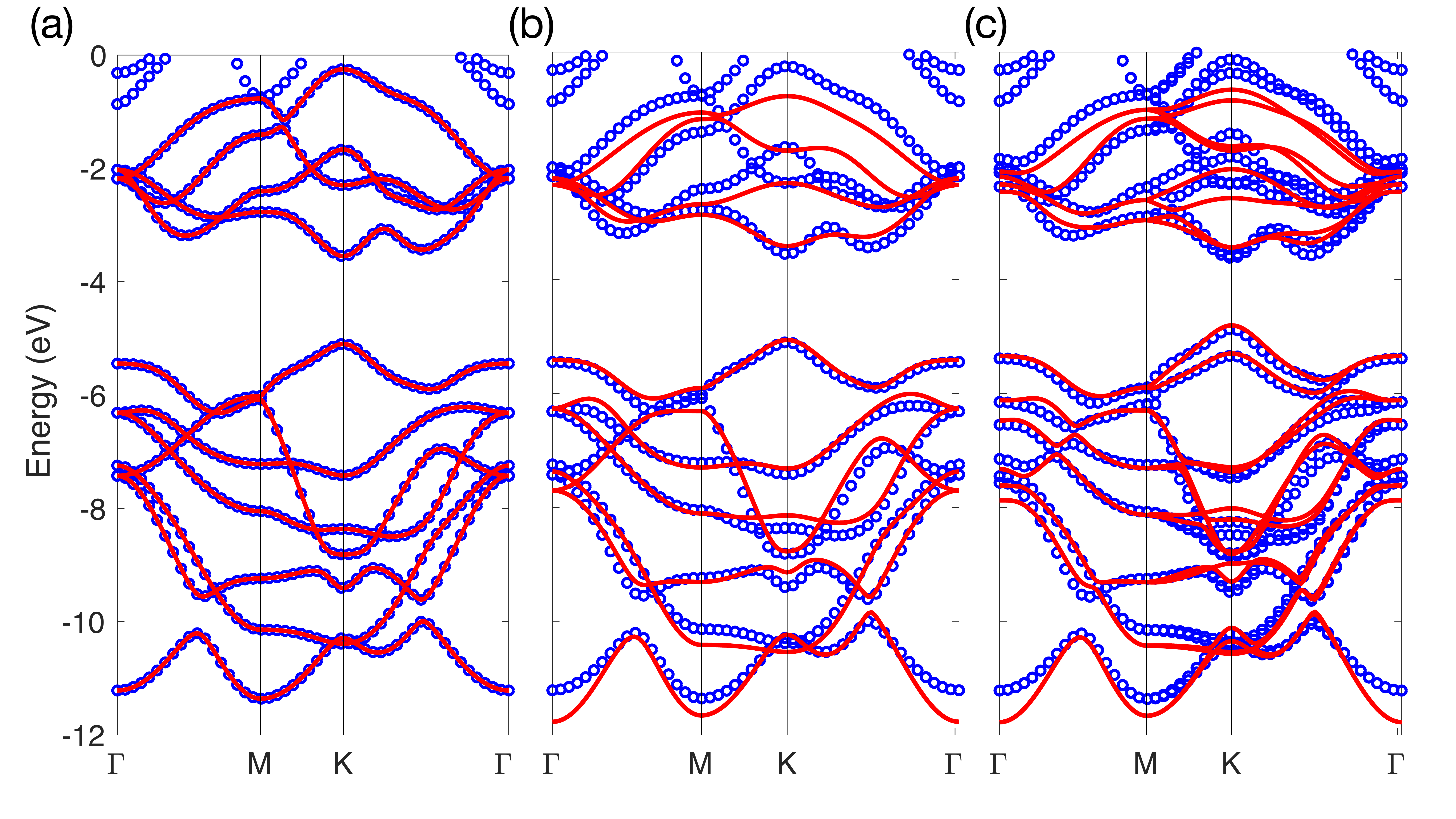}
\caption{Wannier tight-binding Hamiltonian construction from DFT for the WSe$_2$ monolayer: (a) DFT band structure, blue circles (without spin-orbit coupling) with the eleven $p-d$ hybrid bands which are relevant for low-energy electronic properties, used to derive the Wannier tight-binding Hamiltonian (red lines). (b) Wannier Hamiltonian results with truncation to limit the range of neighbor coupling terms. (c) Hamiltonian augmented by the atomic spin-orbit coupling terms (red lines), compared with the full DFT calculation with spin-orbit coupling included (blue circles).}
\label{fig:wannier_construction}
\end{figure}

\end{widetext}

The slowly varying in-plane strain field can be described by the displacement deformation vector field $\vec{u}=(u_x(x,y),u_y(x,y))$. The coordinates $x$ and $y$ denote the undistorted crystal coordinate, which is mapped to the new position $(x+u_x(x,y),y+u_y(x,y))$ in space. Since a constant displacement field introduces no physical changes to the layers, the strain field is characterized by the derivative of $\vec{u}$, defined in tensor form

\begin{equation}
u_{ij} = \frac{1}{2}(\partial_i u_j + \partial_j u_i)
\end{equation} with $i,j$=$x,y$. This 2nd-rank tensor can be decomposed into the trace scalar part $u_{xx}+u_{yy}$, and the doublet ($u_{xx}-u_{yy}$, $-2u_{xy}$) which forms a two-dimensional irreducible representation of the $C_{3v}$ symmetry group of the crystal. There is also a rotational piece, $\omega_{xy}=\partial_x u_y-\partial_y u_x$ which we take $\omega_{xy}=0$ by choosing the proper set of coordinates. We can further simplify the modeling by applying the local density approximation to the strain effects, that is, by assuming locally the tight-binding parameters are approximated by the strained periodic crystal with constant $u_{ij}$. In the following, these strain model parameters are extracted from the Wannier transformation of DFT calculations with periodic unit cells for the uniformly strained crystals. A structure with non-uniform strain can be modeled by combining these local-strain tight-binding parameters which have only long-wavelength variations compared to the lattice constants.

The key steps in constructing these microscopic Hamiltonians are:

(i) In linear elastic theory, the deformed microscopic displacement vector $\vec{v}'=(v_x',v_y',v_z')$ between atomic sites is

\begin{equation}
\label{cauchy-born}
\begin{split}
&v_x'=v_x+ v_x \partial_x u_x + v_y \partial_y u_x \\
&v_y'=v_y+ v_x \partial_x u_y + v_y \partial_y u_y \\
&v_z'=v_z 
\end{split}
\end{equation} with $\vec{v}=(v_x,v_y,v_z)$ the unstrained vector. Though these relations hold for the primitive lattice vectors, strictly speaking this approximation, the Cauchy-Born rule\cite{Cauchy_Born_1,Cauchy_Born_2}, is only valid for a Bravais lattice with a single atom basis. For a strained primitive unit cell with multiple basis atoms, the relative position or orientation of these atoms varies, in addition to the relations prescribed by Eq. (\ref{cauchy-born}). For example, in layered materials such as phosphorene, TMDCs and puckered graphene-like materials, there is a height variation in the position of individual atoms under strain. We adopt the approximation of Eq. (\ref{cauchy-born}) in modeling graphene and hBN for simplicity. We include the height corrections for the chalcogen atoms in TMDCs by generalizing the above Cauchy-Born approximation.

(ii) To incorporate the strain effects in the tight-binding Hamiltonians, the $t^0_{\alpha \beta}$ hopping integral between $\alpha,\beta$ orbitals on different sites is assumed to scale with the pair distance $|\vec{\delta}_{\alpha \beta}|$, known as the central force approximation. Up to leading order linear response, the strained hopping integral can be approximated as\cite{qshe_tmdc}

\begin{equation}
\label{eqn:central_force}
t'_{\alpha\beta}=t^0_{\alpha\beta} + \mu \vec{\delta}_{\alpha \beta} \cdot (\vec{\delta}_{\alpha \beta} \cdot \vec{\bigtriangledown}) \cdot \vec{u},
\mu=\frac{1}{|\delta_{\alpha \beta}|}[\frac{dt_{\alpha\beta}}{d|\delta_{\alpha\beta}|}]
\end{equation} Some empirical models go beyond the linear order by proposing a functional form which depends on the pair distance, such as exponential functions\cite{graphene_rev,graphene_unistrain_tb} or algebraic functions of $|r|$.\cite{wills_harrison} For the orbitals that are not $s$-like, the hopping integrals within the two-center Slater-Koster approximation\cite{slater} can be decomposed into various channels related to the angular momentum projection, such as the $\sigma$ and $\pi$ bonds in $p$-$p$ orbital coupling. The scaling can be applied to each channel as a function of pair distance. In general, the scaling of the hopping integral reflects the shapes of the orbitals and the changes in the crystal field potential. These translate into more involved forms of scaling beyond merely the pair distance dependence. For example, if the crystal is stretched along a direction that is perpendicular to the bond, the central force approximation would dictate no change for the hopping, which is not accurate. Here, we derive the models up to linear order in the strain and beyond the central force approximation. All the terms that couple $(u_{xx}+u_{yy})$, $(u_{xx}-u_{yy})$ and $u_{xy}$ are retained in the Hamiltonian, and their forms are constrained by the underlying crystal symmetry. Thus, the hoppings along a bond acquire corrections when the crystal is stretched along the perpendicular direction to the bond, which captures the local environment change. Many layered materials involve orbitals beyond $s$-like ones, and have a more complicated geometry for atomic configurations and relative orientations. 

(iii) Treatments of strain effects on tight-binding Hamiltonians typically involve only the scaling of hopping terms and neglect the variations for on-site energy terms. The on-site energy variations will be relevant for a layer with non-uniform strain field, also called the deformation potential. We extract the relevant potential information and work function from DFT calculations and define the energy reference point to be zero at the vacuum level outside the layer. In experiments, the presence of a substrate or encapsulating layers, and the charge redistribution in the layer with non-uniform strain result in further modification of the electrostatic environment, the screening for interactions and hence of the onsite terms. Solving the self-consistent potential profile is beyond the scope of the current treatment. 

(iv) To complete our discussion in the presence of the macroscopic perpendicular (out-of-plane) displacements $h$ for the layer or the flexural phonon mode in the long wavelength, we can define the generalized strain tensor\cite{novel_strain_review}

\begin{equation}
\tilde{u}_{ij} = \frac{1}{2}(\partial_i u_j + \partial_j u_i+\partial_i h \partial_j h)
\end{equation}  We expect $\tilde{u}_{ij}$ to capture part of the contributions to the strained tight-binding Hamiltonians. Due to (mirror) symmetry breaking and curvature effects, other terms with derivatives of $h$ that couple states of different sectors can also appear, which can lead to interesting phenomena such as spin-lattice couplings in layered materials\cite{soc_flexural1,soc_flexural2,tmdc_strain_curv}. Capturing these contributions require a Wannier transformation to extract parameters for a curved layer in a supercell geometry, which we leave for future work.


\subsection{Application to Monolayer Graphene and hBN}
In graphene, the semi-metallic gapless $p_z$ bands feature relativistic linear Dirac dispersion at low-energy near the K points of the BZ. Most of the electronic properties can be explained by the simple two-band model involving only the $p_z$ orbitals. hBN can be viewed as a closely related structure to graphene, with a gapped insulating band structure introduced by the Semenoff mass terms\cite{Semenoff_mass} from the sublattice symmetry breaking. For the monolayer modeling of strained graphene and hBN, the distorted atomic positions at the $A/B$ basis sites are assumed to follow Eq. (\ref{cauchy-born}). In terms of electronic modeling, we retain only $p_z$ orbitals up to third nearest neighbor coupling. This is adequate to give a very good description of the key features of the band structure, especially at the band extrema\cite{graphene_tbh}. To model the strain effects for graphene and hBN, we first start with the on-site potential energy term, which is defined relative to the DFT vacuum level outside the layer and can be written as

\begin{equation}
\label{eqn:ghbn_onsite}
\epsilon = \epsilon_0 +\alpha_0 (u_{xx}+u_{yy})
\end{equation} to leading order in $u_{ij}$. The linear coupling to the two-dimensional representation ($u_{xx}-u_{yy},-2u_{xy}$) is forbidden from the underlying crystal and $p_z$ orbital symmetry. For the near-neighbor hopping terms, the strain-dependent tight-binding parameters can be written as

\begin{equation}
\label{eqn:ghbn_hop}
t_{\vec{r}} = t_{\vec{r}}^0+ \alpha_{\vec{r}} (u_{xx}+u_{yy}) + \beta_{\vec{r}}[\omega_y^{\vec{r}}(u_{xx}-u_{yy})+2\omega_x^{\vec{r}}u_{xy}]
\end{equation} where $\vec{r}$ is the bond vector, $\hat{\omega}^{\vec{r}}=(\omega^{\vec{r}}_x,\omega^{\vec{r}}_y)$ ($|\hat{\omega}|=1$) is the associated unit vector, and $\alpha_{\vec{r}}$ and $\beta_{\vec{r}}$ are the strain response parameters. The $\hat{\omega}^{\vec{r}}$ unit vector is parallel to the bonding direction for the first and third neighbor hopping, but perpendicular to the second neighbor hopping direction (see Eq. (\ref{eq_graphene_1NN}) for the first neighbor example). This form is constrained by the irreducible representation of the underlying crystal symmetry. The central force approximation would further constrain the $\alpha_{\vec{r}}$ and $\beta_{\vec{r}}$ parameters. For example, the nearest neighbor terms under this approximation would have $\alpha_1=-\beta_1$, which clearly is not sufficient as our detailed modeling shows.

For graphene and hBN, the relevant parameters that enter Eq. (\ref{eqn:ghbn_hop}) are tabulated in Table \ref{tab:ghbn} with unit vector defined as $\hat{\omega}_\theta=\cos(\theta)\hat{x}+\sin(\theta)\hat{y}$. In this table, only the independent hopping terms along specific directions as shown in Fig. \ref{fig:crystal_convention} (a) are tabulated. The rest of the bonds at the equivalent positions can be related by appropriate symmetry operations.

\begin{widetext}

\begin{table}[h!]
  \centering
  \caption{On-site, Eq. (\ref{eqn:ghbn_onsite}), and nearest neighbor hopping parameters, Eq. (\ref{eqn:ghbn_hop}), for graphene and hBN. For hBN, the superscript indicates the starting point of the hopping matrix element (otherwise from $A$ site to $B$ site.). The vector $\vec{\delta}=(\vec{a}_1+2\vec{a}_2)/3$ and the units are in eV. The last column specifies the corresponding $\hat{\omega}^{\vec{r}}$ unit vectors as in Eq. (\ref{eqn:ghbn_hop}).}
  \label{tab:ghbn}
  \begin{tabular}{cllll}
    Graphene & & &\\
    \hline
    on-site & $\epsilon_0^{\rm C}=-3.613$ & $\alpha_0^{\rm C}=-4.878$ & \\
   $\vec{\delta}$  & $t^0_1=-2.822$ & $\alpha_1=4.007$ & $\beta_1=-3.087$ & $\hat{\omega}_{\pi/2}$ \\
   $\vec{a}_1$ &  $t^0_2=0.254$ & $\alpha_2=-0.463$ & $\beta_2=0.802$ & $\hat{\omega}_{\pi/2}$\\
   $\vec{\delta}-\vec{a}_1-2\vec{a}_2$ &  $t^0_3=-0.180$ & $\alpha_3=0.624$ & $\beta_3=0.479$ & $\hat{\omega}_{-\pi/2}$\\
    \hline
    hBN & & &\\
    \hline
    On-site   & $\epsilon^{\rm B}_0=-1.287$ & $\alpha^{\rm B}_0=-4.778$ & \\
     & $\epsilon^{\rm N}_0=-5.393$ &$\alpha^{\rm N}_0=-2.227$ \\
     $\vec{\delta}$   & $t^0_1=-2.683$ & $\alpha_1=3.142$ & $\beta_1=-2.386$ & $\hat{\omega}_{\pi/2}$ \\
    $\vec{a}_1$ & $t_2^{\rm 0B}=0.048$ & $\alpha_2^{\rm B}=0.176$ & $ \beta_2^{\rm B}=1.061$ & $\hat{\omega}_{\pi/2}$\\
    $\vec{a}_1$ & $t_2^{\rm 0N}=0.218$ & $\alpha_2^{\rm N}=-0.231$ & $\beta_2^{\rm N}=0.721$ & $\hat{\omega}_{\pi/2}$ \\
    $\vec{\delta}-\vec{a}_1-2\vec{a}_2$ & $t^0_3=-0.228$ & $\alpha_3=0.419$ & $\beta_3=0.598$ & $\hat{\omega}_{-\pi/2}$\\
    \hline
  \end{tabular}
\end{table}

\end{widetext}

\subsection{Application to Transition Metal Dichalcogenides}

The monolayer TMDCs with H structure are semiconductors with a direct band gap (typically 1-2 eV), with band structures that have similar features to those of hBN (an insulator) with the band edges at the K valleys. We start our formulation with the tight-binding Hamiltonian in the monolayer TMDC crystal. The relevant states consist of seven valence bands and four conduction bands, which are hybrids of metal $d$ orbitals and chalcogen $p$ orbitals. In Fig. \ref{fig:wannier_construction} we illustrate the DFT (blue circles) and Wannier construction for WSe$_2$ monolayer crystal with the tight-binding bands for these $p-d$ orbital hybrids in red lines. The $xy$ layer mirror symmetry can be utilized to classify these states into odd and even sectors, with the band edges being in the even sector. We focus on the spinless models and group the odd/even orbitals as $\Psi_A=(\phi_x=d_{xz}^{(o)}, \phi_y=d_{yz}^{(o)}, -)$, $\Psi_B=(\phi_x=p_{x}^{(o)}, \phi_y=p_{y}^{(o)}, \phi_z=p_{z}^{(o)})$, $\Psi_C=(\phi_x=d_{xy}^{(e)}, \phi_y=d_{x^2-y^2}^{(e)}, \phi_z=d_{z^2}^{(e)})$ and $\Psi_D=(\phi_x=p_{x}^{(e)}, \phi_y=p_{y}^{(e)}, \phi_z=p_{z}^{(e)})$ with the $xy$ mirror plane still being a symmetry of the crystal when in-plane strain is included, the $(o/e)$ superscript denoting the odd/even sector (the $z$ component is omitted in the $\Psi_A$ group). The grouping and the $\phi_x$, $\phi_y$ and $\phi_z$ labelings are related to the $x$-, $y$-, $z$-like orbitals under three-fold rotation symmetry of the crystal. For the Hamiltonians below, we will classify coupling terms between different groups of orbitals as

\begin{equation}
\langle \Psi_i | H | \Psi'_j\rangle=\begin{bmatrix}
    H_{xx} & H_{xy} & H_{xz} \\
    H_{yx} & H_{yy} & H_{yz} \\
    H_{zx} & H_{zy} & H_{zz} \\
\end{bmatrix}
\end{equation}  where $H_{\alpha\beta}=\langle \phi_\alpha^i | H | {\phi '}_\beta^{j} \rangle$. For the strained TMDC monolayer crystal, the local optimum atomic configurations show that the distance $d_{\rm X-X}$ for the chalcogen pair varies as

\begin{equation}
\label{eqn:TMDC_xx}
\frac{1}{2}d_{\rm X-X}=d_0-d_1(u_{xx}+u_{yy})
\end{equation} with the form constrained by the three-fold rotation crystal symmetry. The pair distance stretches when the crystal is compressed and the relevant parameters are tabulated in Table \ref{tab:tmdc_x_height}.

\begin{table}[h!]
  \centering
  \caption{The lattice constants (for unstrained TMDCs) $a$ (\AA), and the distance between chalcogen atoms $d_{\rm X-X}$ (\AA) in the strained TMDCs, given by Eq. (\ref{eqn:TMDC_xx}).}
  \label{tab:tmdc_x_height}
  \begin{tabular}{ccccc}
  \hline
   & MoS$_2$ & MoSe$_2$ & WS$_2$ & WSe$_2$ \\
  \hline
  $a$   & $3.182$ & $3.317$ & $3.182$ & $3.316$ \\
  $d_0$ & $1.564$ & $1.669$ & $1.574$ & $1.680$ \\
  $d_1$ & $0.517$ & $0.572$ & $0.560$ & $0.611$ \\
  \hline
  \end{tabular}
\end{table}

In the original tight-binding Hamiltonian of the TMDC crystal\cite{tmdc_tbh}, we included the onsite terms and up to third neighbor couplings. The first and third neighbor couplings are of the M-X type, while the second neighbor is of M-M or X-X type. We investigate the strain correction to these Hamiltonian terms: 

(i) The on-site terms include not only the on-site energy but also hybridization between different orbitals at the same site. The total on-site Hamiltonian has four terms $H_{ii}^{(0)}$ ($i=A, B, C, D$), and they share the same form. Within each sector, this symmetric form is simplified with the three-fold rotation symmetry and the $yz$ mirror symmetry:

\begin{equation}
\label{eqn:TMDC_onsite}
\begin{split}
& \hat{H}^{(0)} = \begin{bmatrix}
\epsilon_1 & 0 & 0 \\
0 & \epsilon_1 & 0 \\
0 & 0 & \epsilon_0
\end{bmatrix}+(u_{xx}+u_{yy})\begin{bmatrix}
\alpha^{\rm (0)}_1 & 0 & 0 \\
0 & \alpha^{\rm (0)}_1 & 0 \\
0 & 0 & \alpha^{\rm (0)}_0
\end{bmatrix} + \\
& (u_{xx}-u_{yy}) \begin{bmatrix}
\beta^{\rm (0)}_0 & 0 & 0\\
0 & -\beta^{\rm (0)}_0 & \beta^{\rm (0)}_1 \\
0 & \beta^{\rm (0)}_1 & 0
\end{bmatrix}  +  2 u_{xy} \begin{bmatrix}
0 & \beta^{\rm (0)}_0 & \beta^{\rm (0)}_1 \\
\beta^{\rm (0)}_0 & 0 & 0 \\
\beta^{\rm (0)}_1 & 0 & 0\\
\end{bmatrix} 
\end{split}
\end{equation} for all four TMDCs.

(ii) First and third neighbor couplings are hoppings from M atoms to X atoms (at $-(\vec{a}_1+2\vec{a}_2)/3$ and $2(\vec{a}_1+2\vec{a}_2)/3$ respectively). There are two groups for the first neighbor coupling $H_{BA}^{(1)}$, $H_{DC}^{(1)}$ and one group for the third neighbor term $H_{DC}^{(3)}$ ($H_{BA}^{(3)}$ is neglected). They all have the following scaling form with strain:

\begin{equation}
\label{eqn:TMDC_first}
\begin{split}
& \hat{H}^{\rm (n)} =\begin{bmatrix}
 t^{\rm (n)}_0 & 0 & 0 \\
0 & t^{\rm (n)}_1 & t^{\rm (n)}_2 \\
0 & t^{\rm (n)}_3 & t^{\rm (n)}_4
\end{bmatrix}+(u_{xx}+u_{yy})\begin{bmatrix}
\alpha^{\rm (n)}_0 & 0 & 0 \\
0 & \alpha^{\rm (n)}_1 & \alpha^{\rm (n)}_2 \\
0 & \alpha^{\rm (n)}_3 & \alpha^{\rm (n)}_4
\end{bmatrix} +\\
& (u_{xx}-u_{yy})\begin{bmatrix}
\beta^{\rm (n)}_0 & 0 & 0\\
0 & \beta^{\rm (n)}_1 & \beta^{\rm (n)}_2 \\
0 & \beta^{\rm (n)}_3 & \beta^{\rm (n)}_4
\end{bmatrix}  +  2u_{xy} \begin{bmatrix}
0 & \beta^{\rm (n)}_5 & \beta^{\rm (n)}_6 \\
\beta^{\rm (n)}_7 & 0 & 0 \\
\beta^{\rm (n)}_8 & 0 & 0\\
\end{bmatrix}
\end{split}
\end{equation} where ${\rm n}=1, 3$ for the first and the third neighbor couplings.


(iii) The second neighbor hoppings are between M-M and X-X pairs (at $\vec{a}_1$ position) and there are four kinds of terms, $H_{ii}^{(2)}$ ($i=A, B, C, D$). They all share the same following form:

\begin{equation}
\label{eqn:TMDC_second}
\begin{split}
& \hat{H}^{(2)} =\begin{bmatrix}
t^{\rm (2)}_0 & t^{\rm (2)}_3 & t^{\rm (2)}_4 \\
-t^{\rm (2)}_3 & t^{\rm (2)}_1 & t^{\rm (2)}_5 \\
-t^{\rm (2)}_4 & t^{\rm (2)}_5 & t^{\rm (2)}_2 \\
\end{bmatrix}+(u_{xx}+u_{yy})\begin{bmatrix}
\alpha^{\rm (2)}_0 & \alpha^{\rm (2)}_3 & \alpha^{\rm (2)}_4 \\
-\alpha^{\rm (2)}_3 & \alpha^{\rm (2)}_1 & \alpha^{\rm (2)}_5 \\
-\alpha^{\rm (2)}_4 & \alpha^{\rm (2)}_5 & \alpha^{\rm (2)}_2 \\
\end{bmatrix} +\\
& (u_{xx}-u_{yy})\begin{bmatrix}
\beta^{\rm (2)}_0 & \beta^{\rm (2)}_3 & \beta^{\rm (2)}_4 \\
-\beta^{\rm (2)}_3 & \beta^{\rm (2)}_1 & \beta^{\rm (2)}_5 \\
-\beta^{\rm (2)}_4 & \beta^{\rm (2)}_5 & \beta^{\rm (2)}_2 \\
\end{bmatrix}  +  2u_{xy} \begin{bmatrix}
0 & \beta^{\rm (2)}_6 & \beta^{\rm (2)}_7 \\
\beta^{\rm (2)}_6 & 0 & \beta^{\rm (2)}_8 \\
\beta^{\rm (2)}_7 & -\beta^{\rm (2)}_8 & 0\\
\end{bmatrix}
\end{split}
\end{equation} 

For convenience, the values of the parameters that enter in the expressions for on-site (superscript 0), first- and third-neighbor (superscript 1 and 3) and second-neighbor (superscript 2) hoppings are collected in a sequence of Tables in the Appendix.

Thus far, we considered only the hopping along one specific direction, which gives the simplest expressions for the hopping matrix elements. The equivalent terms are related to this by the three-fold rotation symmetry or Hermitian conjugation. There are no new independent parameters associated with these terms in the equivalent directions. The form of these hopping directions in the presence of the strain field involves a simple transformation. For example, for the bond $\delta'$ which is rotated counterclockwise by $2\pi/3$  from the bond $\delta$, the hopping is


\begin{equation}
\label{eq_rotation_sym}
\begin{split}
H_{\delta'}(u_{xx},u_{yy}&,2u_{xy}) = \hat{\mathcal{U}}_{\mathcal{R}}^\dagger H_{\delta}(u_{xx}',u_{yy}', 2u_{xy}') \hat{\mathcal{U}}_{\mathcal{R}} \\
u_{xx}' &= u_{xx}/4 +3u_{yy}/4 -\sqrt{3}u_{xy}/2 \\
u_{yy}' &= 3u_{xx}/4 +u_{yy}/4 +\sqrt{3}u_{xy}/2  \\
2 u_{xy}' &= \sqrt{3} u_{xx}/2 -\sqrt{3} u_{yy}/2 -u_{xy} \\
\end{split}
\end{equation} For graphene and hBN, $\mathcal{U}_{\mathcal{R}}=1$. For TMDCs, $H_\delta$ and $H_{\delta'}$ are the 3 $\times$ 3 matrices, as parametrized for the Hamiltonians above, and 

\begin{equation}
\hat{\mathcal{U}}_{\mathcal{R}}=\begin{bmatrix}
-1/2& \sqrt{3}/2 & 0\\
-\sqrt{3}/2 & -1/2 & 0 \\
0 & 0 & 1\\
\end{bmatrix}
\end{equation} with $\hat{\mathcal{U}}_{\mathcal{R}}^3=1$. This three-fold rotation operation together with the Hermitian conjugate which reverse the bond direction complete the parametrization of all equivalent bonds in the tight-binding Hamiltonian.

\begin{widetext}

\begin{figure} [h!]
\centering
\includegraphics[width=0.95\textwidth]{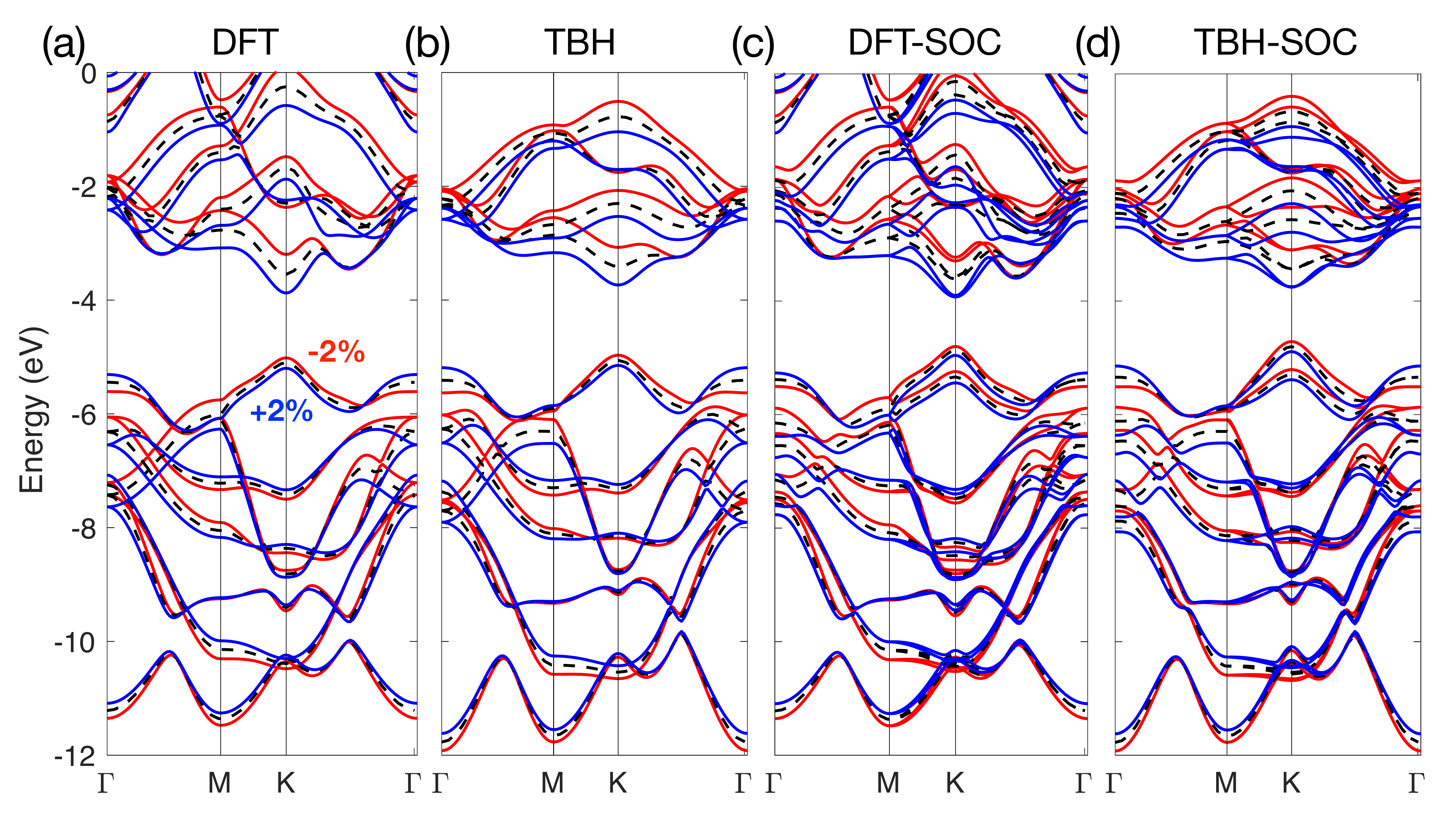}
\caption{Comparison of the (a) DFT and (b) TBH electronic band structure without spin-orbit coupling for a monolayer WSe$_2$ crystal with isotropic strain. The black dashed lines are the bands from the pristine crystal while the red (blue) solid ones are from the crystal with $-2\%$ ($+2\%$) isotropic strain. The high energy bands in the DFT calculations are those beyond the $p-d$ hybrids included in the TBH basis. The vacuum level is at zero energy. (c) and (d), similar comparison with spin-orbit coupling.}
\label{fig:TMDC_STRAIN_DFT_TBH}
\end{figure}

\end{widetext}

For the unstrained TMDC crystal with only $\epsilon_i$ and $t^{(i)}_j$ terms for each interaction, the present model corresponds exactly to the one in our previous work\cite{tmdc_tbh}. The crucial spin-splitting of the bands can be generalized by doubling the orbitals by the spin degrees of freedom and incorporating the spin-orbit coupling as the atomic on-site $\lambda \vec{L} \cdot \vec{S}$ terms\cite{tmdc_tbh}. The symmetry-allowed spin-dependent hopping terms beyond these on-site atomic contributions are neglected in this work, but can be extracted and further modeled based on the Wannier procedure. In Fig. \ref{fig:TMDC_STRAIN_DFT_TBH}, we compare the full DFT calculations as shown in (a) to the simplified TBH in (b) for the pristine WSe$_2$ crystal and the ones with $\pm 2\%$  isotropic strain applied. In (c) and (d) we compare the DFT results with spin-orbit coupling to the TBH augmented with atomic $\lambda \vec{L} \cdot \vec{S}$ on-site terms. We find good agreement between the full DFT calculations and our TBH results. We also note that the couplings to isotropic strain $(u_{xx}+u_{yy})$ have the same form as the unstrained couplings, while the terms with $(u_{xx}-u_{yy})$ and $u_{xy}$ break this form in a pattern that respects the crystal symmetry by forming appropriate symmetry invariants. The previous modeling of the graphene and hBN cases is similar to the $H_{zz}$ terms here; details on the symmetry constraint derivations can be found in the Appendix B.

\begin{figure}
\centering
\includegraphics[width=0.5\textwidth]{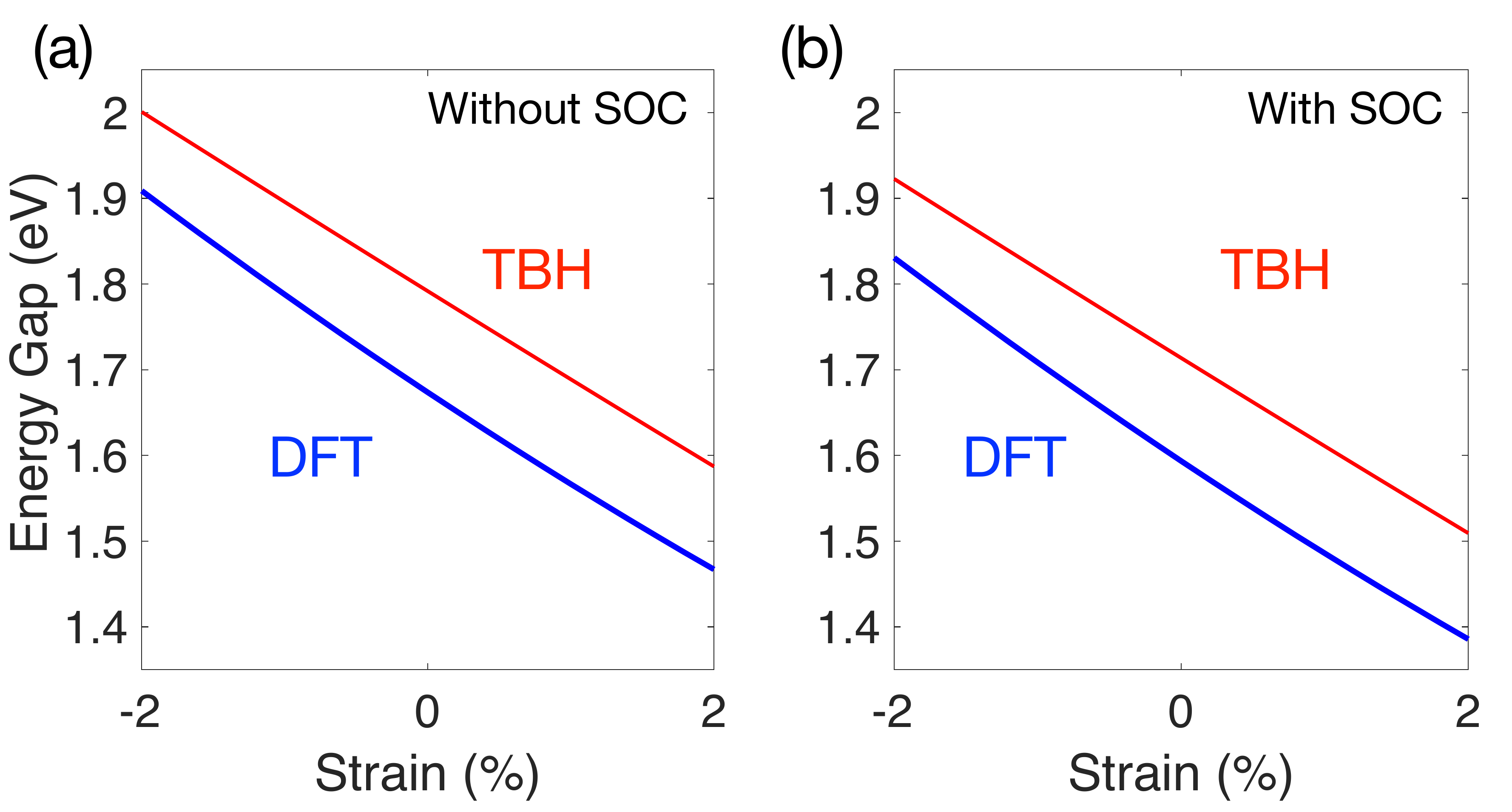}
\caption{The energy gap between the highest valence band and the lowest conduction band at K valley for a MoS$_2$ monolayer crystal under isotropic strain (a) without and (b) with spin-orbit coupling corrections included. The blue (red) lines are from  DFT (TBH) calculations. The slopes agree well for the strain effects.}
\label{fig:SOC_STRAINGAP}
\end{figure}

As a final comment, we discuss some of the important features of the band structure described by our tight-binding Hamiltonian in the presence of strain. The band gap at the K valley scales linearly with the isotropic biaxial strain as shown in Fig. \ref{fig:SOC_STRAINGAP} for MoS$_2$. The slope agrees well between the tight-binding Hamiltonian (red line) which gives -103 meV/\% and the full DFT calculation (blue line) with -110 meV/\%. The relative offset of the two can be corrected by adding more terms of longer range to the truncated Hamiltonian. The slope is also in good agreement with photoluminescence experiments, measured at -105 meV/\% with substrate thermal expansion\cite{mos2_strain_gap} and -99meV/\% with suspended monolayer MoS$_2$\cite{mos2_strain_gap2}. To compare all four TMDCs, recent optical experiments show that MoSe$_2$ < MoS$_2$ < WSe$_2$ < WS$_2$ for the bandgap shifts under biaxial strain\cite{fourTMDC_strain} and the sequence is consistent with our DFT and tight-binding results. The slopes for four TMDCs can be inferred from the $f_4$ parameters in Table \ref{tab:TMDC_KP} for the effective Hamiltonians.

\section{\label{sec:level1}EFFECTIVE HAMILTONIANS}


In this section, we derive the effective Hamiltonian to illustrate the strain effects on the electronic band structure and the symmetry properties, at specific $k$ points relevant to the low energy degrees of freedom.  In the literature, many effective Hamiltonians have been proposed with various levels of accuracy, including the coupling terms to external fields such as strain and electromagnetic fields. One way to arrive at these effective Hamiltonians is through the construction of invariants under the irreducible symmetry group representation\cite{group_theory_md,eff_graphene_strain,eff_graphene_strain}, from objects such as the momentum $\vec{k}$, strain tensor $u_{ij}$ and other fields present. Though symmetry group analysis alone cannot determine the numerical coupling parameters, it is useful to identify all the symmetry allowed terms in the effective theory. An alternative way of deriving the effective Hamiltonians is based on the expansion of the tight-binding Hamiltonians for the material\cite{tmdc_tb_strain}. The additional irrelevant high energy bands at the expansion $k$ point can be integrated out\cite{sch_wolff} and various effective terms can be generated in the reduced space of the low energy bands. The order of expansion can be controlled and the numerical coupling constants can be derived from the tight-binding Hamiltonian parameters. In the following, we rederive the lowest order effective Hamiltonians and show that they are consistent with the ones in the literature, which is a cross-check of the symmetry properties of our tight-binding Hamiltonians. Higher order effective terms can be generated by further expanding the model\cite{tmdc_tb_strain}.

\subsection{Monolayer Graphene}
For the single layer graphene, the electronic band structure exhibits linear gapless Dirac cones at two inequivalent K$_\pm$ points. Around the K$_+$ point, we define the wavefunction as $\Psi_{k}^+ = (\Psi_{(K_++k)}^A,\Psi_{(K_++k)}^B)$ for the components on the $A/B$ sublattice at momentum $(K_++k)$, and $\hat{\sigma}$ the Pauli matrices on sublattice indices. The three nearest $B$ sites from the central $A$ site are located at

\begin{equation}
\delta_1^{(1)}=\frac{a}{\sqrt{3}}(0,1), \delta_2^{(1)}=\frac{a}{\sqrt{3}}(\frac{-\sqrt{3}}{2},\frac{-1}{2}), \delta_3^{(1)}=\frac{a}{\sqrt{3}}(\frac{\sqrt{3}}{2},\frac{-1}{2})
\end{equation} with $a$ the lattice constant. Under uniform strain, the changes in the hopping strength from $A$ to $B$ sites are

\begin{equation}
\label{eq_graphene_1NN}
\begin{split}
\delta t_1^{(1)}&= \alpha_1 (u_{xx}+u_{yy}) +\beta_1 (u_{xx}-u_{yy}) \\
\delta t_2^{(1)}&= \alpha_1 (u_{xx}+u_{yy}) -\beta_1 (u_{xx}-u_{yy})/2 - \sqrt{3} \beta_1 u_{xy} \\
\delta t_3^{(1)}&= \alpha_1 (u_{xx}+u_{yy}) -\beta_1 (u_{xx}-u_{yy})/2 + \sqrt{3} \beta_1 u_{xy}
\end{split}
\end{equation} using the transformation rule of Eq. (\ref{eq_rotation_sym}). The same procedure applies to the second and third neighbors. Together with the on-site terms, we arrive at the k $\cdot$ p Hamiltonian after expanding the tight-binding Hamiltonian $H_{TB}(\vec{k})$ at K$_+$.

\begin{equation}
H_{K_+} = v_F H_0(\vec{k}) + a_0' H_0' + \sum_{i=1}^{5} a_i H_i(\vec{k})
\end{equation} with $\vec{k}=(k_x,k_y)$ the momentum measured from K$_+$ and the definition for each term and the coefficients are given in Table \ref{tab:graphene_kp}. $H_0$ gives the usual Dirac Hamiltonian with linear dispersion with $H_0'+H_1$ the shift in Dirac energy from the on-site and second nearest neighbor contributions. The $H_i$ terms with $i>0$ are the strain induced contributions\cite{eff_graphene_strain}. $H_2$ is the pseudo gauge field term which shifts the Dirac point. In the non-uniformly strained crystal, this term will depend on the spatial position and is responsible for generating pseudo Landau levels. A term $H_6=[\partial_y(u_{xx}-u_{yy})+2\partial_x u_{xy}]\hat{\sigma}_z$ implies a gap-opening in the presence of non-uniform strain field\cite{eff_graphene_strain} which can be estimated from the changes of on-site terms in the uniform strain field.

\begin{widetext}

\begin{table}[h!]
  \centering
  \caption{Effective low-energy Hamiltonians at K$_+$ valley including strain terms for graphene. $\hat{\sigma}_x$, $\hat{\sigma}_y$ are Pauli matrices, with length $l=a/\sqrt{3}$ where $a$ is the graphene lattice constant. The numerical values are from the upper part of Table \ref{tab:ghbn} for graphene with units of energy.} 
  \label{tab:graphene_kp}
  \begin{tabular}{ccc}
    \hline
    $H_0$ & $\hat{\sigma}_x k_x l +\hat{\sigma}_y k_y l$ & $-\frac{3}{2}t^0_1+3t^0_3$\\
    $H_0'$ &  $\mathbb{1}$ & $\epsilon^{\rm C}_0-3t^0_2$  \\
    $H_1$ & $(u_{xx}+u_{yy})\mathbb{1}$ & $\alpha^{\rm C}_0-3\alpha_2$\\
    $H_2$ & $(u_{xx}-u_{yy})\hat{\sigma}_x-2u_{xy}\hat{\sigma}_y$ & $\frac{3}{2}(\beta_1-\beta_3)$ \\
    $H_3$ & $[(u_{xx}-u_{yy})k_x l-2u_{xy}k_y l]\mathbb{1}$ & $\frac{9}{2}\beta_2$\\
    $H_4$ & $(u_{xx}+u_{yy})(\hat{\sigma}_x k_x +\hat{\sigma}_y k_y)$ & $-\frac{3}{2}(\alpha_1+\frac{\beta_1}{2}-2\alpha_3+\beta_3)$ \\
    $H_5$ & $u_{ij} \hat{\sigma}_i k_j l; i,j=x,y$ & $\frac{3}{2}\beta_1+3\beta_3$ \\
    \hline
  \end{tabular}
\end{table}

\end{widetext}


\subsection{Transition Metal Dichalcogenides}
The spinless TMDC tight-binding Hamiltonian consists of eleven bands. We project the full model to the reduced two-band model at the K$_+$ point, consisting of the highest valence band $\Phi^v$ (of $d_{x^2-y^2}+{\rm i} d_{xy}$ character) and the lowest conduction band $\Phi^c$ (of $d_{z^2}$ character) and we investigate the effects of uniform $u_{ij}$ strain field. The spin-orbit coupling can be incorporated with additional spin dependent terms. The full Hamiltonian is $H_{\rm TB}(\vec{k})=H^0_{\rm TB}(\vec{k})+H_{\rm strain}$. To derive the leading order effective two-band Hamiltonian, we expand the unstrained $H^0_{\rm TB}(\vec{k})$ to linear order in $\vec{k}$ and take the strain part $H_{\rm strain}$ to be proportional to the strain field $u_{ij}$ without additional $k$ dependence. The reduced band effective Hamiltonian can be determined by the matrix elements $H^{\rm eff}_{i,j}=\langle \Phi^i | H_{\rm TB}(\vec{k}) | \Phi^j \rangle, (i,j)=(c,v)$, labeling the valence $(v)$ and conduction $(c)$ bands. We choose the convention $\Phi_{k}^+ = (\Phi_{(K_++k)}^c,\Phi_{(K_++k)}^v)$, with $\hat{\sigma}$ acting upon those two band indices. For the unstrained TMDC, the effective k $\cdot$ p Hamiltonian takes the form of a massive Dirac fermion\cite{qshe_tmdc}:

\begin{equation}
H^{0}_{\rm TB}=f_0 \mathbb{1}+\frac{f_1}{2}\hat{\sigma}_z+f_2 a(k_x \hat{\sigma}_x+k_y \hat{\sigma}_y)
\end{equation} with $a$ the lattice constant (see Table \ref{tab:tmdc_x_height}), $f_0$ the midgap position relative to the vacuum level, $f_1$ the mass gap term and $f_2$ the velocity in the Dirac equation. The spin splitting can be captured by adding $(1\pm\hat{\sigma}_z) \hat{s}_z$ terms with $\hat{s}$ the Pauli matrices on the spin indices of the enlarged spin-band Hilbert space. For the lowest order correction terms in the presence of deformations\cite{qshe_tmdc}, the additional strain terms in the Hamiltonian are

\begin{equation}
\begin{split}
H_{\rm strain}=&f_3 \sum_i u_{ii} + f_4 \sum_i u_{ii} \hat{\sigma}_z \\
&+ f_5 [(u_{xx}-u_{yy})\hat{\sigma}_x-2u_{xy} \hat{\sigma}_y]
\end{split}
\end{equation} with $f_3$ ($f_4$) modifying the midgap position (massive gap), and $f_5$ the pseudo gauge field term. Each term contributes a symmetry invariant term by the appropriate product of various objects\cite{qshe_tmdc}. The values of the parameters for all four TMDCs are given in Table \ref{tab:TMDC_KP} based on the expansion of the tight-binding Hamiltonian. For the higher-order corrections in $\vec{k}$ and the strain field $u_{ij}$ in this reduced band Hamiltonian, there are two kinds of terms that contribute: the ones from the direct expansion of the full Hamiltonian within the subspace, and the virtual coupling process to higher levels via Schrieffer-Wolff transformation\cite{sch_wolff,tmdc_tb_strain}.


When an out-of-plane deformation and curvature are present in the layer, the mirror symmetry is broken and the odd/even states can mix. With the spin-orbit couplings taken into account, various types of coupling terms will be generated which relate spin, band, strain field and curvature\cite{soc_flexural1,soc_flexural2}, and have been shown to introduce a spin-lattice coupling as an in-plane effective magnetic field in the TMDC lattice\cite{tmdc_strain_curv}.


\begin{table}[h!]
  \centering
  \caption{TMDC k $\cdot$ p theory parameters at K with the units in eV.}
  \label{tab:TMDC_KP}
  \begin{tabular}{ccccccc}
    TBH & $f_0$ & $f_1$ &  $f_2$ & $f_3$ & $f_4$ &  $f_5$ \\
    \hline
    MoS$_2$ & $-5.07$ & $1.79$ & $1.06$ & $-5.47$ & $-2.59$ & $2.20$ \\ 
    MoSe$_2$ & $-4.59$ & $1.55$ & $0.88$  & $-5.01$ & $-2.28$ & $1.84$ \\
    WS$_2$ & $-4.66$ & $1.95$ & $1.22$ & $-5.82$ & $-3.59$ & $2.27$ \\
    WSe$_2$ & $-4.23$ & $1.65$ & $1.02$ & $-5.26$ & $-3.02$ & $2.03$ \\
	\hline
  \end{tabular}
\end{table}




\section{\label{sec:level1}CONCLUSION}
We constructed {\it ab initio} tight-binding models for the strained layered materials using Wannier transformation of DFT calculations which bridges the microscopic tight-binding Hamiltonians and the effective Hamiltonians based on symmetry principles, using graphene, hBN, and TMDCs as prototypical examples. This method is free from any empirical fitting procedures and captures the microscopic details of the electronic coupling to the strain field, or equivalently the long-wavelength in-plane acoustic phonons. These models apply to systems with multiple orbitals of distinct symmetries, going beyond the single scaling Gr\"{u}neisen parameter approach and the central force approximation. Though the linear response regime is assumed throughout the present treatment, aharmonic couplings at larger strain can be included in a similar way. The method can also be generalized to extract the electronic coupling to long wavelength optical phonon modes and the interlayer coupling in the vertically compressed layer stacks.

These microscopic strain models are relevant for a wide range of applications, including: straintronics\cite{straintronics}, that is, engineering the strain field to obtain the desired electronic properties such as band gaps and effective masses; the realization of stretchable electronic devices based on the layered materials\cite{graphene_stretch}; exploiting the interplay between moir\'{e} patterns, commensurate-incommensurate transitions\cite{com_incom_GHBN} and distortions\cite{TwBLG_Jahn_Teller} which result from twisted bilayer structure that already strongly modifies the monolayer Dirac dispersion and induces insulating states from the superlattice\cite{TwBLG_exp}; exploring the effects of topological lattice defects\cite{graphene_rev,top_defect}; induced interference effects from lattice deformation\cite{graphene_ab}; understanding of electronic scattering and mobility from lattice deformations. The pseudo magnetic field, that does not break time-reversal symmetry, induced by the strain field may be utilized to probe many-body physics through the quantum oscillations without magnetic field\cite{qosc}, or fractional Josephson effect when coupled with a superconductor\cite{strain_graphene_josephson}. Beyond the applications involving static strain fields, we also expect that our microscopic analysis is applicable to the dynamical strain field generated by oscillating acoustic waves\cite{graphene_acoustic}, which can be used as an experimental probe of other excitations in materials, or as a means to realize periodically modulated Floquet Hamiltonians, which will be relevant for studies of non-equilibrium or topological phases\cite{Floquet}. This extracted electron-phonon coupling is also relevant to understand the Raman spectroscopy\cite{graphene_raman,graphene_raman_eph} and other phonon-mediated phenomena.



\begin{acknowledgements}
We thank Rodrick Kuate Defo, Jiayu Shen, Jhih-Shih You, Yen-Hung Ho, Dennis Huang, Daniel Massatt, Paul Cazeaux, Mitchell Luskin, Francisco Guinea, Bertrand I. Halperin and Philip Kim for useful discussions. This work was supported by the STC Center for Integrated Quantum Materials, NSF Grant No. DMR-1231319 and by ARO MURI Award W911NF-14-0247. M.A.C. acknowledge the support from the Ministry of Science and Technology (Taiwan) under contract number NSC 102-2112-M-007-024-MY5, and Taiwan's National Center of Theoretical Sciences (NCTS). S. F. thanks the National Tsing Hua University for its hospitality. We used the Extreme Science and Engineering Discovery Environment (XSEDE), which is supported by NSF Grant No. ACI-1053575.
\end{acknowledgements}


\begin{appendices}

\section{\label{sec:level1}NUMERICAL METHODS FOR DFT AND WANNIER CONSTRUCTION}
The DFT calculations in this work were carried out using the Vienna Ab initio Simulation Package (VASP)\cite{vasp1,vasp2} with Projector Augmented-Wave (PAW) type of pseudo-potentials, parametrized by Perdew, Burke and Ernzerhof (PBE)\cite{pbe}. A slab geometry with a 20 \AA $ $ vacuum region is used to reduce the interactions between periodic images. The DFT calculations for TMDCs are converged with plane-wave energy cutoff $450$ eV and a reciprocal space grid sampling of size 29 $\times$ 29 $\times$ 1.

The extended Bloch wavefunction basis can be transformed into the maximally-localized Wannier functions (MLWF) basis as implemented in the Wannier90 code\cite{mlwf}. With this transformation, the effective tight-binding Hamiltonian for a designated group of bands of the material can be constructed. This not only gives an efficient numerical method to reproduce DFT results but also provides a physically transparent picture of localized atomic orbitals and their hybridizations. From the calculations with and without spin-orbit coupling, we find that a simple atomic onsite $L \cdot S$ term captures well the full DFT band structure with spin-orbit coupling included. Our work is based on the systematic analysis of such tight-binding Hamiltonians with strain applied in the DFT calculations, which  inherit the {\it ab initio} information without fitting procedures for the numerical parameters. Further corrections for band gaps from advanced GW calculations or other choices of exchange correlation functionals are also compatible with Wannier constructions.

\section{\label{sec:level1}SYMMETRY AND IRREDUCIBLE REPRESENTATIONS}
	
The models presented in this paper can be thought of as a set of linear equations which describe how an operator $\mathcal O$, such as the tight-binding energy between two orbitals or the total mechanical energy, changes under some real-space field $X$, like the strain $u_{ij}$. But even a simple linear model for the next-nearest neighbor hoppings between chalcogen atoms in the TMDCs would be complicated. Such a model is based on how three strain fields affect the hoppings between nine pairs of orbitals ($p_i$ to $p_j$) in three bonding directions, needing a total of $81$ ($3 \times 9 \times 3$) fitting parameters. The number of independent parameters is smaller, as the symmetery operations of the crystal relate the values of some parameters to one another, or require others to be zero. Therefore, when modeling these two-dimensional materials it is vital to understand how the crystal symmetery constrains linear models in order to validate computational results. For example, if one DFT fitted parameter happens to be orders of magnitude smaller than the rest of the parameters, it may be unclear if it should be taken as exactly zero. Performing an analysis of the crystal symmetry can clarify this problem as well as provide some insight into how many calculations would be necessary to create a complete model.
	
One method of understanding the crystal symmetry constraints is through representation theory of finite groups. By knowing what patterns of matrices are compatible with the point group of a given crystal, one can enumerate all possible constrained terms which may arise in the modeling process\cite{eff_graphene_strain}. Another, more practical description of this process is as follows: modify the operator $\mathcal O (X)$  under conjugate symmetry operations of the crystal, denoted as $\mathcal S$; for each $\mathcal S$, one can then generate a set of linear equations by requiring that the physical model remains unchanged under the symmetry, namely 
	
\begin{equation}
\label{eq:sym_op}
\mathcal S^{-1} [ \mathcal O (\mathcal S X \mathcal S^{-1}) ] \mathcal S = \mathcal O (X)
\end{equation} Each $\mathcal S$ does not necessarily generate a unique set of equations, but applying all $\mathcal S$ yields the same constraints as expected from representation theory. An example of this second approach is how lattice strain affects a tensor-valued operator, like the electric field gradient (EFG)\cite{hBN_EFG}. However, for the tight-binding energies, the approach is not so obvious. Whereas scalar or tensor-valued quantities can be directly written as finite-dimensional representations of the point group of the crystal, the Wannier localization process can only be considered a representation if the Wannier orbitals themselves obey the crystal symmetries. The localized orbitals must translate, rotate, or reflect into a linear combination of themselves under each crystal symmetry. In practice this does not occur, as the Wannier orbitals are only defined to minimize the spread in the electron density, sometimes breaking crystal symmetry in the process. In our modeling, the Wannier orbitals have small asymmetries, although there are approaches to ensure crystal symmetry exists in the final Wannier orbitals\cite{sym_Wannier}. Assuming they are symmetric allows us to correctly constrain the model, eliminating the numerically introduced asymmetry.
	
First, we consider the nearest-neighbor and 3rd-nearest neighbor hoppings, that is, the ones between TMDC atoms of the same species. Our model includes the $C_3$ rotation symmetry by construction in Eq. (\ref{eq_rotation_sym}), which is simply an implementation of Eq. (\ref{eq:sym_op}) with $\mathcal S$ taken to be rotation by $2\pi/3$. Then we need to consider only the symmetry relating to the chosen bonding direction $t_1$, which is reflection through the $y$-$z$ plane. The Hamiltonian must  be invariant under this symmetry, but the operation affects both the orbitals of the Hamiltonian and the strain field components. Thus, the $xy$ and $xz$ components of the Hamiltonian are constrained to couple only with $u_{xy}$ (both odd under mirror symmetry), while every other component couples only with $u_{xx}$ and $u_{yy}$ (both even under mirror symmetry). For the second nearest neighbor, a similar argument applies. In this case the mirror plane lies halfway between the orbitals, so now we must compare terms in the Hamiltonian to their transpose ($H^{(2)}_{ij}$ and $H^{(2)}_{ji}$). The same rules can be used to check that the 2nd nearest neighbor tight-binding terms are consistent with the mirror symmetry constraint.

Finally, the onsite terms are constrained by the $C_3$ rotation symmetry explicitly. The $u_{xx} + u_{yy}$ correction is diagonal, as it is a 1-dimensional representation, and the ($u_{xx} - u_{yy}$,$-2 u_{xy}$) corrections have their $x$ and $y$ components rotate into one another as a valid 2-dimensional representation.
	
From these considerations we have constructed the form of the tight-binding Hamiltonians given in Eq. (\ref{eqn:TMDC_onsite}) - (\ref{eqn:TMDC_second}). The following tables (\ref{tab:TMDC_AA} through \ref{tab:TMDC_N2}) contain the values of the parameters that enter in the expressions of the model Hamiltonians for the four common TMDCs, namely MoS$_2$, MoSe$_2$, WS$_2$, WSe$_2$.

\begin{widetext}

\begin{table}[h!]
  \centering
  \caption{Onsite ($H_{AA}^{(0)}$) and second neighbor hopping ($H_{AA}^{(2)}$) strain terms in units of eV for MoS$_2$, MoSe$_2$, WS$_2$, WSe$_2$.}
  \label{tab:TMDC_AA}
  \begin{tabular}{c|cccc}
 $H_{AA}^{\rm (n)}$ & MoS$_2$ & MoSe$_2$ & WS$_2$ & WSe$_2$ \\
\hline
\hline
    $\epsilon_1$ & $-4.873$ & $-4.547$ & $-4.327$ & $-4.069$   \\ 
    $\alpha^{\rm (0)}_1$ & $-2.498$ & $-2.341$ & $-2.631$ & $-2.357$   \\ 
    $\beta^{\rm (0)}_0$ & $-0.890$ & $-0.810$ & $-0.986$ & $-0.902$   \\ 
\hline
\hline
    $t^{\rm (2)}_0$ & $-0.206$ & $-0.146$ & $-0.198$ & $-0.137$   \\ 
    $t^{\rm (2)}_1$ & $0.031$ & $0.017$ & $0.027$ & $0.013$   \\ 
    $t^{\rm (2)}_3$ & $-0.257$ & $-0.191$ & $-0.310$ & $-0.232$   \\ 
\hline
    $\alpha^{\rm (2)}_0$ & $-0.258$ & $-0.309$ & $-0.453$ & $-0.490$   \\ 
    $\alpha^{\rm (2)}_1$ & $-0.202$ & $-0.125$ & $-0.213$ & $-0.117$   \\ 
    $\alpha^{\rm (2)}_3$ & $0.705$ & $0.514$ & $0.834$ & $0.589$   \\ 
\hline
    $\beta^{\rm (2)}_0$ & $-0.676$ & $-0.588$ & $-0.942$ & $-0.809$   \\ 
    $\beta^{\rm (2)}_1$ & $-0.192$ & $-0.118$ & $-0.175$ & $-0.090$   \\ 
    $\beta^{\rm (2)}_3$ & $0.555$ & $0.416$ & $0.649$ & $0.480$   \\ 
    $\beta^{\rm (2)}_6$ & $-0.095$ & $-0.063$ & $-0.076$ & $-0.037$   \\     
\hline
\hline
  \end{tabular}
\end{table}

\begin{table}[h!]
  \centering
  \caption{Onsite strain terms ($H_{BB}^{(0)}$, $H_{CC}^{(0)}$, $H_{DD}^{(0)}$) in units of eV for MoS$_2$, MoSe$_2$, WS$_2$, WSe$_2$.}
  \label{tab:TMDC_N0}
  \begin{tabular}{c|ccc|ccc|ccc|ccc}
   & \multicolumn{3}{c|}{ MoS$_2$ } & \multicolumn{3}{c|}{ MoSe$_2$ } & \multicolumn{3}{c|}{ WS$_2$ } & \multicolumn{3}{c}{ WSe$_2$ }\\
\hline
     & $H_{BB}^{(0)}$ & $H_{CC}^{(0)}$ & $H_{DD}^{(0)}$ & $H_{BB}^{(0)}$ & $H_{CC}^{(0)}$ & $H_{DD}^{(0)}$ & $H_{BB}^{(0)}$ & $H_{CC}^{(0)}$ & $H_{DD}^{(0)}$ & $H_{BB}^{(0)}$ & $H_{CC}^{(0)}$ & $H_{DD}^{(0)}$\\
\hline
\hline
        $\epsilon_0$ & $-6.720$ & $-6.082$ & $-8.839$ & $-5.986$ & $-5.559$ & $-8.231$ & $-6.838$ & $-5.734$ & $-9.078$ & $-6.066$ & $-5.267$ & $-8.466$   \\ 
    $\epsilon_1$ & $-7.235$ & $-5.856$ & $-7.850$ & $-6.502$ & $-5.314$ & $-7.110$ & $-7.250$ & $-5.498$ & $-8.033$ & $-6.494$ & $-5.001$ & $-7.277$   \\ 
\hline
    $\alpha^{\rm (0)}_0$ & $1.623$ & $-1.021$ & $-0.858$ & $1.396$ & $-1.090$ & $-0.742$ & $1.743$ & $-1.212$ & $0.158$ & $1.385$ & $-1.012$ & $-0.050$   \\ 
    $\alpha^{\rm (0)}_1$ & $-1.500$ & $-1.817$ & $-3.317$ & $-1.440$ & $-2.023$ & $-3.316$ & $-1.854$ & $-1.916$ & $-4.290$ & $-1.724$ & $-1.967$ & $-4.138$   \\ 
\hline
    $\beta^{\rm (0)}_0$ & $-0.094$ & $-0.370$ & $-1.142$ & $-0.121$ & $-0.296$ & $-1.146$ & $0.089$ & $-0.292$ & $-1.390$ & $0.059$ & $-0.220$ & $-1.337$   \\ 
    $\beta^{\rm (0)}_1$ & $0.273$ & $-0.043$ & $0.720$ & $0.270$ & $0.004$ & $0.829$ & $0.487$ & $0.036$ & $1.586$ & $0.482$ & $-0.022$ & $1.507$ \\ 
    \hline
    \hline
  \end{tabular}
\end{table}

\begin{table}[h!]
  \centering
  \caption{First ($H_{BA}^{(1)}$, $H_{DC}^{(1)}$) and third ($H_{DC}^{(3)}$) neighbor hopping strain terms in units of eV for MoS$_2$, MoSe$_2$, WS$_2$, WSe$_2$.}
  \label{tab:TMDC_N13}
  \begin{tabular}{c|ccc|ccc|ccc|ccc}
   & \multicolumn{3}{c|}{ MoS$_2$ } & \multicolumn{3}{c|}{ MoSe$_2$ } & \multicolumn{3}{c|}{ WS$_2$ } & \multicolumn{3}{c}{ WSe$_2$ }\\
\hline
     & $H_{BA}^{(1)}$ & $H_{DC}^{(1)}$ &  $H_{DC}^{(3)}$  & $H_{BA}^{(1)}$ & $H_{DC}^{(1)}$ &  $H_{DC}^{(3)}$  & $H_{BA}^{(1)}$ & $H_{DC}^{(1)}$ & $H_{DC}^{(3)}$  & $H_{BA}^{(1)}$ & $H_{DC}^{(1)}$ &  $H_{DC}^{(3)}$  \\
\hline
\hline
    $t^{\rm (n)}_0$ & $-0.789$ & $1.411$ & $0.014$ & $-0.695$ & $1.268$ & $0.017$ & $-0.884$ & $1.558$ & $0.010$ & $-0.773$ & $1.399$ & $0.017$   \\ 
    $t^{\rm (n)}_1$ & $2.158$ & $0.652$ & $-0.245$ & $1.941$ & $0.554$ & $-0.215$ & $2.302$ & $0.664$ & $-0.273$ & $2.079$ & $0.567$ & $-0.242$   \\ 
    $t^{\rm (n)}_2$ & - & $-0.940$ & $-0.150$ & - & $-0.874$ & $-0.155$ & - & $-0.993$ & $-0.154$ & - & $-0.905$ & $-0.161$   \\ 
    $t^{\rm (n)}_3$ & $-1.379$ & $-0.954$ & $-0.221$ & $-1.326$ & $-0.858$ & $-0.223$ & $-1.436$ & $-0.943$ & $-0.265$ & $-1.401$ & $-0.853$ & $-0.263$   \\ 
    $t^{\rm (n)}_4$ & - & $-0.883$ & $-0.069$ & - & $-0.772$ & $-0.069$ & - & $-1.005$ & $-0.066$ & - & $-0.896$ & $-0.068$   \\ 
\hline
    $\alpha^{\rm (n)}_0$ & $0.545$ & $-0.486$ & $0.173$ & $0.408$ & $-0.407$ & $0.175$ & $0.585$ & $-0.609$ & $0.537$ & $0.406$ & $-0.493$ & $0.468$   \\ 
    $\alpha^{\rm (n)}_1$ & $-0.605$ & $0.843$ & $0.204$ & $-0.417$ & $0.825$ & $0.185$ & $-0.482$ & $1.045$ & $0.185$ & $-0.322$ & $0.917$ & $0.202$   \\ 
    $\alpha^{\rm (n)}_2$ & - & $2.178$ & $0.567$ & - & $1.928$ & $0.554$ & - & $2.827$ & $0.623$ & - & $2.409$ & $0.653$   \\ 
    $\alpha^{\rm (n)}_3$ & $1.845$ & $0.446$ & $0.744$ & $1.718$ & $0.272$ & $0.760$ & $1.826$ & $0.071$ & $1.055$ & $1.764$ & $0.022$ & $1.050$   \\ 
    $\alpha^{\rm (n)}_4$ & - & $-0.208$ & $0.035$ & - & $-0.298$ & $0.062$ & - & $-0.241$ & $-0.090$ & - & $-0.238$ & $-0.021$   \\ 
\hline
    $\beta^{\rm (n)}_0$ & $-1.076$ & $1.724$ & $-0.178$ & $-0.897$ & $1.530$ & $-0.164$ & $-1.128$ & $2.402$ & $-0.345$ & $-0.929$ & $1.973$ & $-0.321$   \\ 
    $\beta^{\rm (n)}_1$ & $0.401$ & $-0.353$ & $-1.069$ & $0.264$ & $-0.367$ & $-0.995$ & $0.140$ & $-0.900$ & $-1.110$ & $-0.029$ & $-0.877$ & $-1.094$   \\ 
    $\beta^{\rm (n)}_2$ & - & $-2.204$ & $-0.070$ & - & $-1.995$ & $-0.093$ & - & $-2.293$ & $-0.125$ & - & $-2.153$ & $-0.114$   \\ 
    $\beta^{\rm (n)}_3$ & $-2.100$ & $-0.682$ & $-0.267$ & $-1.874$ & $-0.510$ & $-0.292$ & $-1.990$ & $-0.306$ & $-0.120$ & $-1.879$ & $-0.276$ & $-0.241$   \\ 
    $\beta^{\rm (n)}_4$ & - & $-0.850$ & $-0.281$ & - & $-0.727$ & $-0.290$ & - & $-1.184$ & $-0.536$ & - & $-0.897$ & $-0.476$   \\ 
    $\beta^{\rm (n)}_5$ & $0.859$ & $0.899$ & $-0.690$ & $0.770$ & $0.761$ & $-0.664$ & $0.915$ & $0.902$ & $-1.093$ & $0.798$ & $0.761$ & $-1.022$   \\ 
    $\beta^{\rm (n)}_6$ & - & $-0.542$ & $-0.382$ & - & $-0.475$ & $-0.391$ & - & $-0.193$ & $-0.644$ & - & $-0.300$ & $-0.651$   \\ 
    $\beta^{\rm (n)}_7$ & $-0.377$ & $-2.093$ & $-0.340$ & $-0.469$ & $-1.841$ & $-0.299$ & $-0.634$ & $-2.934$ & $-0.535$ & $-0.690$ & $-2.447$ & $-0.423$   \\ 
    $\beta^{\rm (n)}_8$ & $-0.836$ & $1.101$ & $0.015$ & $-0.717$ & $1.005$ & $0.007$ & $-0.944$ & $1.427$ & $-0.127$ & $-0.793$ & $1.082$ & $-0.058$   \\ 
    \hline
    \hline
  \end{tabular}
\end{table}

\begin{table}[h!]
  \centering
  \caption{Second neighbor hopping ($H_{BB}^{(2)}$, $H_{CC}^{(2)}$, $H_{DD}^{(2)}$) strain terms in units of eV for MoS$_2$, MoSe$_2$, WS$_2$, WSe$_2$.}
  \label{tab:TMDC_N2}
  \begin{tabular}{c|ccc|ccc|ccc|ccc}
   & \multicolumn{3}{c|}{ MoS$_2$ } & \multicolumn{3}{c|}{ MoSe$_2$ } & \multicolumn{3}{c|}{ WS$_2$ } & \multicolumn{3}{c}{ WSe$_2$ }\\
\hline
     & $H_{BB}^{(2)}$ & $H_{CC}^{(2)}$ & $H_{DD}^{(2)}$ & $H_{BB}^{(2)}$ & $H_{CC}^{(2)}$ & $H_{DD}^{(2)}$ & $H_{BB}^{(2)}$ & $H_{CC}^{(2)}$ & $H_{DD}^{(2)}$ & $H_{BB}^{(2)}$ & $H_{CC}^{(2)}$ & $H_{DD}^{(2)}$\\
\hline
\hline
    $t^{\rm (2)}_0$ & $0.865$ & $0.275$ & $0.912$ & $0.964$ & $0.251$ & $0.991$ & $0.873$ & $0.355$ & $0.965$ & $0.977$ & $0.320$ & $1.047$   \\ 
    $t^{\rm (2)}_1$ & $-0.187$ & $-0.558$ & $0.006$ & $-0.172$ & $-0.473$ & $-0.004$ & $-0.218$ & $-0.691$ & $0.014$ & $-0.198$ & $-0.584$ & $0.003$   \\ 
    $t^{\rm (2)}_2$ & $-0.174$ & $-0.298$ & $-0.192$ & $-0.211$ & $-0.264$ & $-0.217$ & $-0.175$ & $-0.371$ & $-0.212$ & $-0.217$ & $-0.333$ & $-0.241$   \\ 
    $t^{\rm (2)}_3$ & $-0.070$ & $-0.249$ & $-0.038$ & $-0.068$ & $-0.201$ & $-0.039$ & $-0.099$ & $-0.304$ & $-0.101$ & $-0.092$ & $-0.245$ & $-0.102$   \\ 
    $t^{\rm (2)}_4$ & $0.100$ & $0.114$ & $-0.106$ & $0.076$ & $0.096$ & $-0.121$ & $0.110$ & $0.145$ & $-0.163$ & $0.079$ & $0.124$ & $-0.185$   \\ 
    $t^{\rm (2)}_5$ & $-0.068$ & $0.410$ & $0.008$ & $-0.074$ & $0.352$ & $0.005$ & $-0.082$ & $0.488$ & $-0.031$ & $-0.091$ & $0.423$ & $-0.038$   \\ 
\hline
    $\alpha^{\rm (2)}_0$ & $-1.841$ & $-1.027$ & $-1.425$ & $-1.979$ & $-0.951$ & $-1.586$ & $-1.844$ & $-1.232$ & $-1.122$ & $-1.986$ & $-1.127$ & $-1.357$   \\ 
    $\alpha^{\rm (2)}_1$ & $-0.027$ & $1.544$ & $-0.057$ & $-0.103$ & $1.333$ & $-0.072$ & $-0.067$ & $1.947$ & $-0.162$ & $-0.152$ & $1.617$ & $-0.159$   \\ 
    $\alpha^{\rm (2)}_2$ & $0.444$ & $1.032$ & $0.644$ & $0.536$ & $0.885$ & $0.668$ & $0.434$ & $1.123$ & $0.674$ & $0.557$ & $1.013$ & $0.718$   \\ 
    $\alpha^{\rm (2)}_3$ & $-0.045$ & $0.206$ & $-0.170$ & $-0.059$ & $0.195$ & $-0.162$ & $-0.042$ & $0.462$ & $-0.314$ & $-0.074$ & $0.325$ & $-0.303$   \\ 
    $\alpha^{\rm (2)}_4$ & $-0.210$ & $0.285$ & $-0.199$ & $-0.123$ & $0.236$ & $-0.202$ & $-0.208$ & $0.365$ & $-0.333$ & $-0.105$ & $0.291$ & $-0.287$   \\ 
    $\alpha^{\rm (2)}_5$ & $0.141$ & $-0.738$ & $0.065$ & $0.142$ & $-0.596$ & $0.050$ & $0.177$ & $-0.654$ & $0.105$ & $0.188$ & $-0.564$ & $0.112$   \\ 
\hline
    $\beta^{\rm (2)}_0$ & $-2.203$ & $-0.910$ & $-2.013$ & $-2.378$ & $-0.793$ & $-2.180$ & $-2.254$ & $-1.068$ & $-1.920$ & $-2.427$ & $-0.966$ & $-2.086$   \\ 
    $\beta^{\rm (2)}_1$ & $0.768$ & $1.337$ & $0.828$ & $0.827$ & $1.108$ & $0.884$ & $0.772$ & $1.240$ & $1.039$ & $0.834$ & $1.179$ & $1.069$   \\ 
    $\beta^{\rm (2)}_2$ & $0.350$ & $0.376$ & $0.540$ & $0.445$ & $0.333$ & $0.576$ & $0.283$ & $0.522$ & $0.580$ & $0.401$ & $0.406$ & $0.556$   \\ 
    $\beta^{\rm (2)}_3$ & $-0.065$ & $-0.003$ & $0.143$ & $-0.016$ & $0.008$ & $0.155$ & $-0.054$ & $-0.083$ & $0.345$ & $0.015$ & $-0.044$ & $0.331$   \\ 
    $\beta^{\rm (2)}_4$ & $-0.208$ & $0.188$ & $-0.056$ & $-0.146$ & $0.126$ & $-0.026$ & $-0.198$ & $0.179$ & $0.062$ & $-0.104$ & $0.129$ & $0.063$   \\ 
    $\beta^{\rm (2)}_5$ & $0.096$ & $-0.779$ & $0.082$ & $0.112$ & $-0.667$ & $0.073$ & $0.127$ & $-0.863$ & $0.130$ & $0.152$ & $-0.727$ & $0.112$   \\ 
    $\beta^{\rm (2)}_6$ & $0.482$ & $-0.634$ & $0.744$ & $0.567$ & $-0.565$ & $0.777$ & $0.467$ & $-0.960$ & $0.858$ & $0.550$ & $-0.776$ & $0.873$   \\ 
    $\beta^{\rm (2)}_7$ & $-0.146$ & $0.288$ & $0.051$ & $-0.128$ & $0.255$ & $0.066$ & $-0.128$ & $0.484$ & $0.146$ & $-0.157$ & $0.308$ & $0.109$   \\ 
    $\beta^{\rm (2)}_8$ & $-0.089$ & $-0.152$ & $-0.099$ & $-0.092$ & $-0.110$ & $-0.127$ & $-0.117$ & $-0.046$ & $-0.236$ & $-0.129$ & $-0.099$ & $-0.224$   \\ 
    \hline
    \hline
  \end{tabular}
\end{table}

\end{widetext}

\clearpage

\end{appendices}

\bibliography{2Dbibref}

\end{document}